\documentclass[12pt]{article}  
\usepackage{amssymb}
\usepackage{amsmath,amsthm}
\theoremstyle{plain}
\newtheorem{thm}{Theorem}[section]
\newtheorem{prop}[thm]{Proposition}
\newtheorem{cor}[thm]{Corollary}
\newtheorem{lemma}[thm]{Lemma}

\theoremstyle{definition}

\newtheorem{rmk}[thm]{Remark}
\newtheorem{rmks}[thm]{Remarks}

\newtheorem{prob}{Problem}

\newcommand{\lra}{\longrightarrow}
\newcommand{\NN}{\mathbb{N}}
\newcommand{\Z}{\mathbb{Z}}
\newcommand{\FF}{\mathbb{F}}
\newcommand{\M}{\mathcal{M}}
\newcommand{\OO}{\mathcal O}

\newcommand{\ver}{\vspace*{-1.5mm}}
\newcommand{\eqr}[1]{~\mbox{$(${\rm \ref{#1}}$)$}}
\newcommand{\mat}{\operatorname{Mat}}

\title{Cryptanalysis of the CFVZ cryptosystem\thanks{Research
    supported by Swiss National Science Foundation Grant no.\ 
    107887 and by Generalitat Valenciana grant CTESPP-2005-060.
    The research of this paper was accepted for presentation at
    the 10th Rhine Workshop on Computer Algebra (RWCA) to be held
    in Basel, Switzerland March 16, 17.}}
\author{Joan-Josep Climent\\
  {\normalsize Departament de Ci\`{e}ncia de la Computaci\'{o}}\ver\\
  {\normalsize i Intel{$\cdot$}lig\`{e}ncia Artificial}\ver\\
  {\normalsize Universitat d'Alacant}\ver\\
  {\normalsize Campus de Sant Vicent del Raspeig}\ver\\
  {\normalsize E-03080 Alacant, Spain}\ver\\
  {\normalsize jcliment@dccia.ua.es}\ver \and
  Elisa Gorla\\
  {\normalsize Department of Mathematics}\ver\\
  {\normalsize University of Z\"urich}\ver\\
  {\normalsize Winterthurerstr 190}\ver\\
  {\normalsize CH-8057 Z\"urich,  Switzerland}\ver\\
  {\normalsize http://www.math.unizh.ch/aa/} \and
  Joachim Rosenthal\\
  {\normalsize Department of Mathematics}\ver\\
  {\normalsize University of Z\"urich}\ver\\
  {\normalsize Winterthurerstr 190}\ver\\
  {\normalsize CH-8057 Z\"urich,  Switzerland}\ver\\
  {\normalsize http://www.math.unizh.ch/aa/}}

\begin{document}

\maketitle

\begin{abstract}
  The paper analyzes a new public key cryptosystem whose
  security is based on a matrix version of the discrete logarithm
  problem over an elliptic curve. 

  It is shown that the complexity of solving the underlying
  problem for the proposed system is dominated by the complexity
  of solving a fixed number of discrete logarithm problems in the
  group of an elliptic curve. Using an adapted Pollard rho
  algorithm it is shown that this problem is essentially as hard
  as solving one discrete logarithm problem in the group of an
  elliptic curve.
\end{abstract}

\medskip


\newpage

\section{Introduction}

Public-key cryptography, based on the intractability of the
discrete logarithm problem, was introduced by Diffie and
Hellman~\cite{di76}.  The Diffie-Hellman protocol allows two
parties Alice and Bob, who are communicating over an insecure
channel, to generate a shared secret key which is difficult to
compute for an eavesdropper.

The discrete logarithm problem (DLP) over various finite groups
has been studied extensively. In the early days the main example
has been the multiplicative group over a finite field $\FF_q$.
Odoni, Varadharajan and Sanders~\cite{od84} introduced the
discrete logarithm problem for matrices over $\FF_{q}$ and a
Diffie-Hellman key exchange protocol based on matrices.  However,
Menezes and Wu~\cite{me97a} reduced the discrete logarithm
problem for matrices to some discrete logarithm problems over
small extensions of $\FF_{q}$.

In the late eighties Miller~\cite{mi86} and Koblitz~\cite{ko87a}
independently proposed to study the DLP in the group of
$\FF_q$-rational points of an elliptic
curve. This was the start of an active research in the area of
elliptic curve cryptography (ECC), and its use for
implementing public-key protocols such as the Diffie-Hellman key
agreement.  The security of ECC is based on the presumed
intractability of the discrete logarithm problem over the curve.

A vast amount of research has been done on the security and
efficient implementation of ECC. Finite groups based on elliptic
curves are very appealing, as the best algorithms known to tackle
the DLP over an elliptic curve has exponential running time, and
this despite intensive attempts on this problem. The interested
reader may consult the recent book~\cite{co06}.

Recently, Climent, Ferr\'{a}ndez, Vicent and Zamora~\cite{cl05u}
introduced a Diffie-Hellman key exchange protocol which used
a combination of matrix algebra ideas and adding points on an
elliptic curve. We will describe this new cryptosystem CFVZ in the
next section. The main results of this paper will be presented in
Section~\ref{cryptanalysis}. 
We will show that CFVZ can be reduced to the problem
of solving $2rs$ discrete logarithm problems over an elliptic
curve in a simultaneous manner. The complexity for doing this is
considerably less than solving $2rs$ single discrete logarithm
problems over an elliptic curve.

\section{The cryptosystem CFVZ of
  Climent-Ferr\'{a}ndez-Vicent-Zamora}

Let $E$ be an elliptic curve defined over the finite field
$\FF_{q}$, and let $E(\FF_q)$ denote the group of $\FF_q$-rational
points of $E$. Assume that $E(\FF_{q})$ is a cyclic group of order
$n$. Denote by $\mat_{r}(\Z)$ the set of all $r \times r$
matrices with integer entries and denote by $\mat_{r \times
  s}(E(\FF_{q}))$ the set of all $r \times s$ matrices whose
entries are elements of the group $E(\FF_{q})$. Let $r,s$ be fixed
positive integers and consider the set
\[
  \xi
  =
  \left\{ 
    \begin{bmatrix}
      A & \Pi \\
        & B
    \end{bmatrix}
    :
    A \in \mat_{r}(\Z), B \in \mat_{s}(\Z), \Pi \in \mat_{r \times s}(E(\FF_{q}))
  \right\}.
\]

The set $\xi$ is a semigroup with the formal matrix multiplication
\[
  \begin{bmatrix}
    A & \Pi \\
      & B
  \end{bmatrix}
  \begin{bmatrix}
    C & \Phi \\
      & D
  \end{bmatrix}
  = 
  \begin{bmatrix}
    AC & A \Phi + \Pi D \\
       & BD
  \end{bmatrix},
\]
where
\[
  A \Phi 
  = 
  [a_{ij}] [P_{ij}] 
  = 
  [Q_{ij}] 
  \quad \text{with} \quad
  Q_{ij} 
  = 
  \sum_{k=1}^{r} a_{ik} P_{kj}
\]
and similarly for $\Pi D$.

Without loss of generality we will assume that $A$ and $B$ are
matrices defined over $\Z/n\Z$. If $A$ and $B$ are invertible
matrices over the ring $\Z/n\Z$ then we can consider the subgroup
generated by the public element
\[
  \M
  =
  \begin{bmatrix}
    A & \Pi \\
      & B
  \end{bmatrix}.
\]

Let $m \ge 1$ be an integer. A direct computation shows that
$\M^{m}
  =
  \begin{bmatrix}
    A^{m} & \Pi_{m} \\
          & B^{m}
  \end{bmatrix}$
where
\begin{equation} \label{eq1}
  \Pi_{m}
  =
  \sum_{i=0}^{m} A^{m-1-i} \Pi B^{i}.
\end{equation}

One way of setting up a discrete logarithm problem is:
\begin{quote}
  ``Given the matrices $\M$ and $\M^m$, find $m$.''
\end{quote}

As shown in~\cite{cl05u}, the order of $\M$ is the least common
multiple of the orders of $A$ and $B$ and hence the discrete
logarithm problem has the character of a discrete logarithm
problem over the matrix ring.

A more interesting problem was introduced in~\cite{cl05u}, we
will call this problem the 
\begin{quote}
  \textbf{CFVZ discrete logarithm problem:} 
  given $\Pi,\Phi \in \mat_{r \times s}(E(\FF_{q}))$ , find\newline 
  $m \in \Z$ such that $\Phi = \Pi_m$ (whenever such an $m$ exists).
\end{quote}

\begin{rmk}
Notice that if the CFVZ discrete logarithm problem has a solution
$m_0$, then it has infinitely many solutions in $\Z$. In fact,
each element of the coset $m_0+l\Z$ is a solution, if we let $l$
be the order of $\M$. Moreover, it may be $\Pi_m=\Pi_{m_0}$ even
for values of $m$ for which $\M^{m} \neq \M^{m_{0}}$.

Notice in addition that the sequence $\Pi_m$ is obtained from a
recurrence relation, namely
\[
  \Pi_{m} = A \Pi_{m-1} + \Pi B^{m-1}.
\]
In particular, the sequence of the $\Pi_m$ has a period.
However it is not true in general that $\Pi_i=\Pi_j$ implies 
$\Pi_{i+1}=\Pi_{j+1}$.
\end{rmk}

The CFVZ discrete logarithm problem induces a Diffie-Hellman key
exchange in the following way:

\begin{itemize}
\item Alice chooses a private key $k$ and computes
\[
  \M^{k}
  =
  \begin{bmatrix}
      A^{k} & \Pi_{k} \\
            & B^{k}
  \end{bmatrix}.
\]
She takes $\Pi_{k}$ as her public key.
\item Bob chooses a private key $l$ and computes
\[
  \M^{l}
  =
  \begin{bmatrix}
    A^{l} & \Pi_{l} \\
          & B^{l}
  \end{bmatrix}.
\]
He takes $\Pi_{l}$ as his public key.
\item Then Alice and Bob consider matrices
\[
  \mathcal{R}
  =
  \begin{bmatrix}
    A & \Pi_{l} \\
      & B
  \end{bmatrix}
  \quad \text{and} \quad
  \mathcal{S}
  =
  \begin{bmatrix}
    A & \Pi_{k} \\
      & B
  \end{bmatrix}
\]
respectively and compute
\[
  \mathcal{R}^{k}
  =
  \begin{bmatrix}
    A^{k} & (\Pi_{l})_{k} \\
          & B^{k}
  \end{bmatrix}
  \quad \text{and} \quad 
  \mathcal{S}^{l}
  =
  \begin{bmatrix}
    A^{l} & (\Pi_{k})_{l} \\
          & B^{l}
  \end{bmatrix}
\]
respectively.  
\end{itemize}
The shared secret is then by equation (\ref{eq1})
\[
  (\Pi_{l})_{k}
  =
  \sum_{j=0}^k  A^{k-1-j} 
  \left(
    \sum_{i=0}^{l} A^{l-1-i} \Pi B^{i}
  \right)
  B^{j}
  =
  \sum_{i=0}^l  A^{l-1-j} 
  \left(
    \sum_{j=0}^{k} A^{k-1-j} \Pi B^{j}
  \right)
  B^{i}
  =
  (\Pi_{k})_{l},
\]
which both Alice and Bob can readily compute.

In order to attack the cryptosystem the following Diffie-Hellman
problem has to be solved:

\begin{prob}\label{DHP1}
Given the matrix $\M$, and the two public keys
$\Pi_k$ and $\Pi_l$, find $(\Pi_k)_l=(\Pi_l)_k$.
\end{prob}


\section{Cryptanalysis of the system}
\label{cryptanalysis}

In this section we analyze the security of the CFVZ
Diffie-Hellman key exchange as proposed in~\cite{cl05u}. We will
show that solving the Diffie-Hellman Problem has the same
complexity as solving an ECDLP on $E(\FF_q)$ and two linear
system of equations in $2rs$ and $r+s-1$ or fewer
unknowns respectively.

For the applications, the curve $E$ and the field $\FF_q$ are
always chosen so that the group $E(\FF_q)$ has prime order.
However, here we will analyze the case when the group $E(\FF_q)$
is cyclic of order $n$, since this introduces no extra
difficulty.

\subsection{Reduction to a matrix problem}

In a first step we show how to reduce the CFVZ discrete logarithm
problem to a problem involving matrices defined over $\Z/n\Z$ only.
For this assume that $P\in E(\FF_q)$ is a generator of the cyclic
group $E(\FF_q)$.

Let $C = [c_{ij}] \in \mat_{r\times s}(\Z/n\Z)$ be a matrix such that
\[
  CP = \Pi
  \quad \text{where} \quad
  CP = [c_{ij}P].
\]
Define the matrix
\[
  M
  =
  \begin{bmatrix} 
    A & C \\ 
    0 & B 
  \end{bmatrix},
\]
and assume
\[
  M^{k}
  =
  \begin{bmatrix}
    A^{k} & C_{k} \\
          & B^{k}
  \end{bmatrix},
  \quad \text{where} \quad
  C_{k}
  =
  \sum_{i=0}^{k} A^{k-1-i} CB^{i}.
\]
The following lemma is readily verified:

\begin{lemma}
  Let $k$ and $l$ be positive integers and let
  \[
    (C_{l})_{k}
    =
    \sum_{j=0}^k A^{k-1-j} 
    \left(
      \sum_{i=0}^{l} A^{l-1-i} C B^{i}
    \right)
    B^{j}.
  \]
  Then
  \[
    \Pi_k
    =
    C_k P  
    \quad \text{and} \quad
    (\Pi_{l})_{k}
    =
    (C_{l})_{k}P.
  \]
\end{lemma}

Based on this lemma, Problem~\ref{DHP1} is solved if we solve 
a number of discrete logarithm problems over the elliptic
curve $E(\FF_q)$, and the following matrix Diffie-Hellman problem:

\begin{prob}\label{DHP2}
  Given the matrix $M$, and the two public keys $C_k$ and $C_l$,
  find $(C_k)_l=(C_l)_k$.
\end{prob}

In order to solve the CFVZ discrete logarithm problem it is
therefore enough to compute 
\begin{equation} \label{eq4}
  \tau
  :=
  3rs
\end{equation}
discrete logarithm problems over the elliptic curve  $E(\FF_q)$
in order to compute matrices $C_k$, $C_l$ and $C$ such that
\[
  \Pi = C P, 
  \quad 
  \Pi_{k} = C_{k} P, 
  \quad \text{and} \quad
  \Pi_{l} = C_{l} P.
\]
Thereafter one has to tackle the linear algebra
Problem~\ref{DHP2}. 

In the remainder of this subsection we show that solving $\tau$
discrete logarithm problems over the elliptic curve $E(\FF_q)$
with regard to a fixed generator $P$ is considerably less complex
than solving $\tau$ individual discrete logarithm problems.
We now analyze the complexity of solving a fixed number of DLPs
in a given cyclic group. We also refer the reader to~\cite{ku01p}
for a treatment of the same problem.

For this assume that $P_1,\ldots,P_\tau$ are points on the
elliptic curve group $E(\FF_q)$. We would like to find integers
$n_1,\ldots,n_\tau$ such that:
\[
  P_i = n_i P, 
  \quad \text{for} \quad
  i=1,\ldots,\tau.
\]
Using an adapted version of the Pollard rho algorithm we
compute points of the form:
\[
  Q_j
  =
  \sum_{i=1}^\tau c_{ij}P_i + d_j P
  \quad \text{with} \quad
  c_{ij}, d_j \in \Z/n\Z.
\]
We repeat this computation until there are more than $\tau$ equal
pairs $Q_i=Q_j$ and $i\neq j$. This is a generalized birthday
problem.  Let $I_{ij}$ be the random variable having the value
$1$ if $Q_i=Q_j$ and the value zero otherwise and consider the
random variable
\[
  W := \sum_{i<j} I_{ij}.
\]
We are interested that
\begin{equation} \label{eq2}
  \mathbb{P}(W\geq \tau)
  >
  \frac{1}{2}
\end{equation}
where $\tau$ is defined by (\ref{eq4}). 
As explained in~\cite[p. 104-107]{ba92b}
(compare also with the recent survey~\cite{da05}) the random
variable $W$ is well approximated by a Poisson random variable.
Based on this fact, the probability of expression (\ref{eq2}) 
can be computed in the following way:

Assume that $\alpha$ points $Q_j$ were computed. Let
\begin{equation} \label{eq3}
  \lambda := \binom{\alpha}{2}/n.
\end{equation}
Then the probability in (\ref{eq2}) is approximated by the expression:
\[
  \mathbb{P}(W\geq \tau)
  =
  1-\sum_{i=0}^{\tau-1} \frac{\lambda^i}{i!}e^{-\lambda}.
\]

Already in the early 18'th century de Moivre~\cite[p.~214]{ha90}
was interested in the maximal value $\tau$ such that
$\mathbb{P}(W\geq \tau)\geq\frac{1}{2}$. Equivalently we can seek
the minimal value $\alpha$ such that with probability more than
$1/2$ there will be at least $\tau$ collisions.

Viewing the Poisson distribution as the limit of a binomial
distribution with expected value $\lambda$ given by (\ref{eq3}), 
one readily gets the approximation 
\[
  \tau
  \leq
  \binom{\alpha}{2}/n,
\]
or equivalently
\[
  \sqrt{\alpha(\alpha-1)} \geq \sqrt{2\tau n}.
\]
The expected number of point additions for the $\tau$
discrete logarithm problems over $E(\FF_q)$ is therefore
$\OO(\sqrt{rsn})$. 

Once we have $t\geq \tau$ collisions we immediately obtain a
system of $t$ linear equations:
\[
  T
  \begin{bmatrix}
    P_1 \\ 
    \vdots \\ 
    P_\tau 
  \end{bmatrix}
  =
  \begin{bmatrix}
    v_1 \\ 
    \vdots \\ 
    v_\tau 
  \end{bmatrix}
  P 
  =
  v P,
\]
where $T \in \mat_{t\times \tau}(\Z/n\Z)$ and the vector $v \in
(\Z/n\Z)^\tau$. As soon as $T$ has full rank $\tau$, the points
$P_i$ can all be computed from $P$ through a simple matrix
inversion of $T$. The cost of inverting $T$ over $\Z/n\Z$
requires $\OO(\tau^3)$ modular multiplications.

In order to simultaneously solve the given $\tau$ discrete
logarithm problems, we can also follow a different approach. Let
$d$ be the determinant of the matrix 
$T \in \mat_{\tau\times  \tau}(\Z/n\Z)$ 
that we obtain after collecting $\tau$ relations
among the given points. Let $g=\gcd(d,n)$ be the greatest common
divisor of $d$ and $n$, and let $m=n/g$. Then $T$ has full rank
over the ring $\Z/m\Z$. Hence a simple matrix inversion gives us
$a_1,\ldots,a_{\tau}\in\Z/m\Z$ such that $n_i=a_i$ modulo $m$ for
all $i=1,\ldots,\tau$.  Because of the algorithm of Pohlig and
Hellman, for all practical purposes we can assume that $n$ is of
the form $n=lp$, where $p$ is prime and $l$ is small. The
probability that the determinant $d$ is invertible modulo $p$ is
equal to
\[
  \frac{|GL_{\tau}(\Z/p\Z)|}{|\mat_{\tau\times\tau}(\Z/p\Z)|}
  =
  \prod_{i=1}^{\tau}\left(1-\frac{1}{p^i}\right).
\]
Here $|GL_{\tau}(\Z/p\Z)|$ denotes the number 
of invertible matrices of size $\tau\times \tau$ over $\Z/p\Z$, 
$|\mat_{\tau\times\tau}(\Z/p\Z)|$ denotes the number of
$\tau \times \tau$ matrices over $\Z/p\Z$. Therefore, with high
probability we can determine the value of $n_1,\ldots,n_{\tau}$
modulo $p$. If $l$ is small, then it is feasible to compute the 
$\tau \left[l/2\right]$ points $a_iP,
(a_i+p)P,\ldots,(a_i+(\left[l/2\right]-1)p)P$ for
$i=1,\ldots,\tau$, where $\left[l/2\right]:=\min\{b\in\Z\;|\;
2b\geq l\}$. Comparing them with $P_i$ and $-P_i$ one can recover the
value of $n_i$ modulo $n$.

If $r$ and $s$ are chosen relatively small in comparison to the size
$n$ of the elliptic curve, then the computation of the matrices
$C_k$, $C_l$ and $C$ is dominated by the task to find at least
$3rs$ collisions, and this task has an expected complexity of
$\OO(\sqrt{rsn})$ point additions.

\subsection{Solution of the matrix problem}

We are giving the matrix $M$ in block-form, with $A \in
\mat_{r\times r}(\Z/n\Z)$, $C \in \mat_{r\times s}(\Z/n\Z)$, and
$B \in \mat_{s\times s}(\Z/n\Z)$. We are working under the
assumption that both $A$ and $B$ are invertible. In fact, as we
will see in the sequel we do not need this assumption in the
analysis of the complexity of Problem~\ref{DHP2}.

We can regard the operation of associating $C_i$ to $C$ as a map
\[
  \begin{array}{rcl}
    -_i : \mat_{r\times s}(\Z/n\Z) & \lra    & \mat_{r\times s}(\Z/n\Z) \\
      C                            & \mapsto & C_i
   \end{array}.
\]
The next lemma shows that the map distributes with respect to the
sum.

\begin{lemma}\label{dist+}
  For any $U, V \in \mat_{r\times s}(\Z/n\Z)$ we have the identity
  \[
    (U+V)_i = U_i+V_i
    \quad \text{for} \quad
    i\in \mathbb{N}.
  \]
\end{lemma}
\begin{proof}
Let 
\[
  M_X
  =
  \begin{bmatrix} 
    A & X \\ 
    0 & B
  \end{bmatrix}
\]
for $X = U, V, U+V$.  
Then $X_i$ is defined by
\[
  (M_X)^i
  =
  \begin{bmatrix} 
    A^i & X_i \\ 
    0 & B^i
  \end{bmatrix},
\]
hence $X_i = A X_{i-1} + X B^{i-1}$. 
We prove the thesis by induction on $i$. If $i=1$, then
\[
  (U+V)_1 
  = 
  U+V 
  = 
  U_1+V_1
\]
and the thesis is readily verified.
Assume that $(U+V)_{i-1}=U_{i-1}+V_{i-1}$ and prove the
analogous identity for $i$. We have
\begin{eqnarray*}
  (U+V)_i
    & = & A(U+V)_{i-1} + (U+V)B^{i-1} \\
    & = & AU_{i-1} + AV_{i-1}+UB^{i-1}+VB^{i-1} \\
    & = & U_i + V_i.
\end{eqnarray*}
\end{proof}

In the next lemma we prove that applying the map $-_i$ commutes
with multiplying copies of $A$ on the left, and copies of $B$ on
the right. In fact, the same is true if we multiply on the left
by a matrix that commutes with $A$ and on the right by a matrix
that commutes with $B$.

\begin{lemma}\label{dist*}
  For any $U\in \mat_{r\times s}(\Z/n\Z)$ and for any 
  $j\in\NN$, the following identities hold
  \[
    (A^jU)_i
    =
    A^jU_i,
    \qquad 
    (UB^j)_i = U_iB^j.
  \]
\end{lemma}
\begin{proof}
Let 
\[
  N
  =
  \begin{bmatrix} 
    A & A^j U \\ 
    0 & B
  \end{bmatrix},
\]
then $(A^jU)_i$ is defined by
\[
  N^i
  =
  \begin{bmatrix} 
    A^i & (A^j U)_i \\ 
    0   & B^i
  \end{bmatrix}.
\]
We prove the thesis by induction on $i$. 
If $i=1$ then $(A^j U)_1 = A^j U = A^j U_1$, so the thesis is true.  
Assume that $(A^j U)_{i-1} = A^jU_{i-1}$ and prove the analogous 
identity for $i$. 
By direct computation, using the induction hypothesis, we obtain
\begin{eqnarray*}
  (A^j U)_i
    & = & A (A^j U)_{i-1} + (A^j U)B^{i-1} \\
    & = & A (A^{j} U_{i-1}) + A^j (U B^{i-1}) \\
    & = & A^j (A U_{i-1} + U B^{i-1}) \\
    & = & A^jU_i.
\end{eqnarray*}

We can obtain the second identity by a similar argument.
\end{proof}

In the next proposition we show how Problem~\ref{DHP2} can be
reduced to solving a linear system over $\Z/n\Z$.

\begin{prop}\label{main}
  Consider the linear system 
  \begin{equation}\label{syst}
    C_k = a_1 C_1 + \cdots + a_{r+s-1} C_{r+s-1}
  \end{equation} 
  where $C_1,\ldots,C_{r+s-1},C_k \in \mat_{r\times s}(\Z/n\Z)$ are
  known, and $a_1,\ldots,a_{r+s-1}\in\Z/n\Z$ are the unknowns.
  The system has (at least) a solution. Any solution of
  (\ref{syst}) determines a homogeneous linear form
  $f_k(x_1,\ldots,x_{r+s-1})=a_1x_1+\cdots+a_{r+s-1}x_{r+s-1} \in
  (\Z/n\Z)[x_1,\ldots,x_{r+s-1}]$ such that for all $l\in\NN$ one
  has
  \[
    (C_l)_k
    =
    f_k(C_l,(C_l)_2,\ldots,(C_l)_{r+s-1}).
  \]
\end{prop}
\begin{proof}
Let $\chi_{M}(x) = \det(xI-M)$ be the characteristic polynomial
of $M$. Since $\chi_{M}(M)=0$, then there exist
$\alpha_0,\ldots,\alpha_{r+s-1} \in \FF_p$ such that
\[
  M^k
  =
  \sum_{i=0}^{r+s-1}\alpha_i M^i.
\]
Hence by definition
\[
  C_k
  =
  \sum_{i=0}^{r+s-1}\alpha_i C_i
  =
  \sum_{i=1}^{r+s-1}\alpha_i C_i,
\]
since $C_0=0$. 
Then $(\alpha_1,\ldots,\alpha_{r+s-1})$ is a
solution of the linear system (\ref{syst}), in particular the
system always has at least a solution.
  
Now let $(a_1,\ldots,a_{r+s-1})$ be a solution of (\ref{syst}).
We claim that for all $l\in\NN$ one has
\[
  (C_l)_k
  =
  \sum_{i=1}^{r+s-1}a_i (C_l)_i.
\]
The thesis is trivially verified for $l=0$ since $C_0=0$.  If
$l=1$ then $(C_1)_i=C_i$ for all $i$, and
\[
  C_k
  =
  \sum_{i=1}^{r+s-1}a_i C_i
\]
since $(a_1,\ldots,a_{r+s-1})$ is a solution of (\ref{syst}) by
assumption.  We proceed by induction on $l\geq 1$.
  
Assume that the thesis holds for $l-1$ and prove it for $l$. By
induction hypothesis we have that
\[
  (C_{l-1})_k
  =
  \sum_{i=1}^{r+s-1}a_i (C_{l-1})_i.
\]
Since $C_l=AC_{l-1}+C_1B^{l-1}$, then by Lemmas~\ref{dist+} and \ref{dist*} we
have the following chain of equalities
\begin{eqnarray*}
  \sum_{i=1}^{r+s-1} a_i (C_l)_i
    & = & \sum_{i=1}^{r+s-1} 
          a_i \left( AC_{l-1}+C_1B^{l-1} \right)_i \\
    & = & \sum_{i=1}^{r+s-1} a_i (AC_{l-1})_i 
          + 
          \sum_{i=1}^{r+s-1} a_i (C_1B^{l-1})_i \\
    & = & \sum_{i=1}^{r+s-1} a_i A(C_{l-1})_i
          +
          \sum_{i=1}^{r+s-1}a_i(C_1)_iB^{l-1} \\
    & = & A 
          \left[ \sum_{i=1}^{r+s-1} a_i (C_{l-1})_i \right] 
          +
          \left[ \sum_{i=1}^{r+s-1} a_i (C_1)_i \right] 
          B^{l-1} \\
    & = & A (C_{l-1})_k + C_k B^{l-1} \\
    & = & A(C_k)_{l-1} + C_kB^{l-1}
\end{eqnarray*}
where the last equality follows from the fact that for each $i,j$ one
has $(C_i)_j=(C_j)_i$. Moreover, by definition one has that
\[
  A (C_k)_{l-1} + C_k B^{l-1}
  =
  (C_k)_l
  =
  (C_l)_k.
\]
This completes the proof.
\end{proof}

\begin{rmks}
  \mbox{}
  \begin{itemize}
  \item In the proof of Proposition~\ref{main} we do not need to
    make any assumption on the matrices $A,B$. In fact, we only
    require the existence of a polynomial $\chi_{M}(x)$ of degree
    smaller than or equal to $r+s-1$, with the property that
    $\chi_{M}(M)=0$. Such a polynomial $\chi_{M}(x)$ always
    exists, since every square matrix over a finite filed has a
    minimal and characteristic polynomial. In particular, we do
    not need to assume that $A$ and $B$ are invertible.
  \item The system (\ref{syst}) may or may not have a unique
    solution. If the system does not have a unique solution, one of its
    solutions does not necessarily give us enough information to
    recover $A^k$ or $B^k$, hence $k$ (solving a DLP in a matrix
    group). 
  \item The rank of the system (\ref{syst}), hence the dimension
    of the family of solutions of the system itself, is not
    relevant towards the goal of solving Problem~\ref{DHP2}. In
    fact, it follows from Proposition~\ref{main} that any
    solution of (\ref{syst}) enables us to compute $(C_l)_k$ from
    the knowledge of $C_k$ and $C_l$. In practice, in order to
    simplify the computations it may be useful to choose a sparse
    solution for the linear system (\ref{syst}) whenever this is
    possible.
  \item A necessary condition for uniqueness of the solution of
    the system (\ref{syst}) is that $M$ be non-derogatory
    (i.e. $\chi_M(x)$ is equal to the minimal polynomial of $M$). 
  \end{itemize}
\end{rmks}

The next corollary is a straightforward consequence of
Proposition~\ref{main}.

\begin{cor}
  With the notation of Section~1 and of Proposition~\ref{main}
  one has
  \[
    (\Pi_l)_k
    =
    f_k(\Pi_l,(\Pi_l)_2,\ldots,(\Pi_l)_{r+s-1}).
  \]
\end{cor}

\section{Complexity Analysis}

In this paper we analyzed the complexity of solving the
Diffie-Hellman Problem, as arising from the Diffie-Hellman
key-exchange proposed in~\cite{cl05u}.  The approach that we
suggest in order to solve the problem is the following:
\begin{enumerate}
\item Use a modified version of the algorithm rho of Pollard and
  find matrices $C, C_k, C_l\in \mat_{r\times s}(\Z/n\Z)$
  such that $C P=\Pi$, $C_k P=\Pi_k$, and $C_l P=\Pi_l$.
\item Compute $C_1,\ldots,C_{r+s-1}$, then find one solution
  $(a_1,\ldots,a_{r+s-1})$ of the linear system
  \begin{equation}  \label{recurrence}
    C_k
    =
    a_1 C_1 + \ldots + a_{r+s-1} C_{r+s-1}.
  \end{equation}
\item Compute $(C_l)_k = a_1(C_l)_1 + \cdots + a_{r+s-1} (C_l)_{r+s-1}$.
\item Compute the secret key $(\Pi_l)_k = (C_l)_k P$.
\end{enumerate}
We showed that the complexity of the first step amounts to
solving $\tau=3rs$ simultaneous DLP's in $E(\FF_q)$ and the
expected complexity is $\OO(\sqrt{rsn})$.

The complexity of the second step amounts to the inversion of a
$(r+s-1)\times (r+s-1)$ matrix over $\Z/ n\Z$. When $n \gg r,s$ this
complexity is polynomial in $\log n$. Similarly the third step is
an easy linear algebra task. Finally the fourth step involves a
number of costly point additions on the elliptic curve.

When $n \gg r,s$ the complexity of the first step dominates the
complexities of the other steps. In this case the complexity of
solving Problem~\ref{DHP1} is at most $\OO(\sqrt{rsn})$.

Instead of computing $3rs$ DPL's it is also possible to only find
the matrices $C$ and $C_k$ by solving $2rs$ DPL's. Like in step~2
one finds $(a_1,\ldots,a_{r+s-1})$ satisfying\eqr{recurrence}.

Using the recurrence relation one then finds
$(\Pi_l)_1,\ldots,(\Pi_l)_{r+s-1}$. From this the secret key
$(\Pi_l)_k$ is readily computed as:
\[
  (\Pi_l)_k
  =
  a_1 (\Pi_l)_1 + \cdots + a_{r+s-1} (\Pi_l)_{r+s-1}.
\]

The advantage of this variant of the algorithm is that only $2rs$
DLP's have to be computed. The disadvantage is that many more
point additions are required in order to compute
$(\Pi_l)_k$. This variant is however faster in situations when
$r,s$ are small in comparison to $n$.

\vspace{.2cm}

\subsection*{Acknowledgments}
The first author would like to thank the University of Z\"urich
for a pleasant stay in Z\"urich which provided the basis of this
research. All authors would like to thank  A.~Barbour for
pointing out Reference~\cite{ha90} and a referee at RWCA for
helpful comments.


\end{document}